\documentclass{appolb}
\usepackage{epsfig}

\begin{document}
\title{ Is early thermalization really needed in A+A collisions?
\thanks{Presented at Workshop on Particle Correlations and Femtoscopy: WPCF-2008}%
}
\author{Yu.M. Sinyukov$^{1}$, A.N. Nazarenko$^{1}$, Iu.A. Karpenko$^{1,2}$
\address{$^{1}$ Bogolyubov Institute for Theoretical
Physics, Metrolohichna str. 14b, 03680 Kiev-143,  Ukraine }
\address{$^{2}$ SUBATECH, University of Nantes – IN2P3/CNRS– EMN, Nantes, France }}
\maketitle

\begin{abstract}
In this note we review our ideas, first published in year 2006,
and corresponding results, including the new ones, which show that
whereas the assumption of (partial) thermalization in relativistic
A+A collisions is really crucial to explain soft physics
observables, the hypotheses of {\it early} thermalization at times
less
than 1 fm/c is not necessary. The reason for the later conclusion is
that the initial transverse flow in thermal matter as well as its
anisotropy, leading to asymmetry of the transverse momentum spectra,
could be developed at pre-thermal, either partonic or classical
field - Glasma, stage with even more efficiency than in the case of
very early perfect hydrodynamics. Such radial and elliptic flows
develop no matter whether a pressure already established. The
general reason for them is an essential finiteness of the system in
transverse direction.
\end{abstract}

\section{Introduction}
The problem of thermalization of the matter in ultra-relativistic
A+A collisions is, certainly, one of the central in this field of
physics. Even the answer to the basic question as for possible
formation of new states of matter, such as the quark-gluon plasma
(QGP), in these experiments depends on a clarification of this
problem.

The experimental data support, in fact, an idea of thermalization.
There are many dynamical models: hydrodynamic one \cite{Landau},
hybrid model - at first hydro evolution, then hadronic cascade
\cite{hybrid}, hydro-kinetic approach - coherent use of
hydrodynamics and kinetics \cite{PRC}, that describe the soft
physics observables in quite satisfactory way. All of them uses an
assumption of thermalization: either complete (local equilibrium),
or partial (with viscosity).

The discovery and theoretical description of elliptic flows at RHIC
\cite{STAR} play especially intriguing role in the problem of
thermalization. It was cleared up \cite{elliptic} that the
anisotropy of transverse momentum spectra, expressed in terms of
$v_2$ coefficients, can be
explained and basically described within hydrodynamic model for
perfect fluid. Since the geometry of initial state in non-central
A+A collision is ellipsoidal-like, where the biggest axis $y$ is
directed orthogonally to the reaction plane, then the collective
velocities will be developed preferably along the smaller axis $x$ -
in reaction plane - since the gradient of pressure is larger in this
direction. It leads to larger blue shift of quanta radiated from
decaying fluid elements in $x$-direction at the freeze-out stage.
This explains the anisotropy of the spectra, but to reach a
quantitative agreement with the experimental data one needs to use
very small initial time, $\tau \sim 0.5$ fm/c, to start the
hydro-evolution. If it starts at later times, neither the collective
velocities, nor their (and the spectra) anisotropy will be developed
enough. These results, in fact, brought the two new ideas: first,
that QGP, at least at the temperatures not much higher $T_c$ , is
the strongly coupled system - sQGP, and so is an almost perfect
fluid, and, second, that thermalization happens at very early times
of collisions.

A necessity for the thermal pressure to be formed as early as
possible appears also in hydrodynamic description of the central
collisions at RHIC. If one starts the hydrodynamic evolution in
"conventional time" $\tau_i=$1 fm/c {\it without} transverse flow,
the latter will not be developed enough to describe simultaneously
the pion, kaon and proton spectra.

The interesting observation related to the same topic was found in
Ref.\cite{Borysova} concerning RHIC HBT puzzle. It was shown that to
describe successfully pion, kaon and proton spectra as well as the
interferometry radii $R_i$ and explain the observed "puzzle",
$R_{out}/R_{side}\approx 1$, one needs i) initial transverse
velocities developed by the time $\tau_i=$1 fm/c and ii) positive
$r_T - t$ correlations between time $t$ and absolute value of
transverse radius $r_T$ on the surface of the particle emission
during the system evolution. In the case of positive correlations
the contribution to $R_{out}^2$, which is proportional to the
duration of the emission squared, can be compensated and, therefore,
$R_{out}/R_{side}$ ratio can be reduced. In Ref.\cite{sin} the
connection between these two conditions, i) and ii), is established.
Namely, in hydrodynamic models the positive $r_T - t$ correlations
at SPS and RHIC energies can appear only if fairly developed
transverse flow already exist at the time $\tau_i=$1 fm/c.

At the first glance, all those results for central collisions just conform
the conclusion coming from non-cental ones: the thermalization and
pressure are established at times essentially smaller than 1 fm/c
and because of this the radial/elliptic flows are already developed
by this time. The crucial problem is, however, that even the most
optimistic theoretical estimates
give thermalization time 1 - 1.5
fm/c \cite{early}, typically, it is 2 - 3 fm/c. The discrepancy
could be even more at LHC energies.

The way to solve the problem was proposed in year 2006, Ref. \cite
{sin}. It was shown that the transverse collective velocities can
be developed at the very early pre-thermal stage with even higher
efficiency than in the locally equilibrated fluid. This result was
developed and exploited in Refs. \cite{Gyul, LHC,
QM}.{\footnote{see also recent contribution to this topic
\cite{Florkowsky, Pratt}.}} In what follows we stick to the main
line developed in these papers.

\section {Early developing of flow and its anisotropy}

A  developing of the initial transverse velocities at the
pre-thermal stage in ultrarelativistic A+A collisions is related to
the complex problem of formation and evolution of chromo- electric
and magnetic fields and their quanta - partons at very early stage
of collision processes. In this note we analyze the developing of
the transverse flows in the two opposite cases: evolution of
non-interacting partons and fields and hydrodynamic expansion of
strongly interacting system - perfect liquid.

\subsection{Non-relativistic analytical results.}

We start from the non-relativistic analytical toy-models. Let us put
the initial momentum distribution of particles with mass $m$ to be
spherically symmetric Gaussian with the width corresponding to the
thermal Boltzmann distribution with uniform temperature $T_{0}$, and
without collective flow: $\mathbf{v}(t=0, \mathbf{r})=0$. Also the initial
spatial distribution of particle density corresponds to the
spherically symmetric Gaussian profile with radius $R_{0}$. Let
particles just to stream flreely. Then, according to \cite{AkkSin}, the
collective velocities, which can be defined at any time $t$
according to Eckart:
\begin{equation}
v^i=\int \frac{d^3p}{m^4}p^if(t,{\bf x};p),
\end{equation}
are
\begin{equation}
\mathbf{v}(t,\mathbf{r})=\mathbf{r}\frac{tT_0}{mR_{0}^{2}+T_0t^2}\rightarrow
\frac{t{\bf r}}{\lambda_{hom}^2}~   {\tt{at}}~  t\rightarrow 0,
\label{flow-sym}
\end{equation}
where $r$ is the radial coordinate and the homogeneity length
$\lambda_{hom}$ is equal  in this case to $R_0$. For the
boost-invariant situation with initially elliptic transverse profile
(corresponding  Gaussian radii are $R_x =\lambda_{hom,x}$ and
$R_y=\lambda_{hom,y}$) \cite{AkkSin} the similar results for
transverse velocities can be obtained at small $t$:
\begin{equation}
v_i(t,\mathbf{r})\sim \frac{tr_i}{\lambda_{hom,i}^2}\label{flow-ell}
\end{equation}

So, the collective velocities,
which are developed in some direction $i$, are inversely proportional
to the corresponding homogeneity length squared,
$\lambda_{hom,i}^2$. The pressure plays no role in this
collisionless process. The collective flow in the expanding system
appears because of a deficit of particles moving inward as compared
to the particles moving outward. It happens due to finiteness
of the system: one has less particles at the periphery than in the
central part (the corresponding deficit is determined by the
homogeneity length), and the flow develops during the evolution
since the momentum of the particles that move outward cannot be
compensated by the momentum of the particles which move inward, even
if at the initial time such a compensation takes place.

It is worth noting that, at least for fairly small time interval,
the flows (\ref{flow-sym}), (\ref{flow-ell}) for collisionless
expansion coincide with flows,
which are developed in hydrodynamically expanding perfect fluid (\cite{AkkSin})
at the same initial conditions.

\subsection{Partonic free-streaming}
Let us start from the initial conditions inspired by the Color Glass
Condensate (CGC) approach. According to the results of Ref.
\cite{KNV} the time $\tau_0 \approx 3/\Lambda_{s}$~(where
$\Lambda_{s}=g^2\mu\equiv 4\pi \alpha_s\mu$ and $\mu^2$ is the
variance of a Gaussian weight over the color charges of partons) is
an  appropriate scale controlling the formation of gluons with a
physically well-defined energy. In Ref. \cite{QM} the results found
in \cite{KNV} for gluon momentum spectra in spatially transverse
homogeneous case are transformed for the transversally finite systems
and longitudinally smeared distribution instead of $\delta(y-\eta)$
($\eta=\frac{1}{2}\ln{\frac{t+x_L}{t-x_L}}\simeq
0,y=\frac{1}{2}\ln{\frac{p_0+p_L}{p_0-p_L}})$. Then the initial
local boost-invariant phase-space density $f$ at  the hypersurface
${(t^2-x_L^2)^{1/2}=\tau_0}$ takes the form \cite{QM}
\begin{eqnarray}\hspace{-25mm}
f(x,p)_{|\tau_0}\equiv
dN/d^3xd^3p_{|\tau_0}=\frac{a_1(b)}{g^2}(\tau_0
m_T\cosh(y-\eta))^{-1}  \nonumber \\
\left(\exp{\left(\sqrt{p_T^2+m_{eff}^2}/T_{eff}\right)}-1\right)^{-1}\frac{\rho(
b,{\bf r}_T)}{\rho_0(b)}, \label{distrib1}
\end{eqnarray}
where $m_{eff}=a_2\Lambda_s$, $T_{eff}=a_3\Lambda_s$; $a_2=0.0358,
a_3=0.465$. For finite systems we approximate the transverse profile
of the gluon distribution by the ellipsoidal Gaussian $\rho(b,{\bf
r}_T)=\rho_0(b) \exp{(-r_x^2/R_x^2-r_y^2/R_y^2)}$ defined from the
best fit of the participant number density in the collisions
with the impact parameter {\bf b}=(b,0). Constant $a_1(b)$ in
Eq.~(\ref{distrib1}) is found then by the comparison of the total
gluon number with results of Ref.~\cite{KNV}.

Let us suppose that an actual thermalization happens about the time
$\tau_i=1 fm/c
>\tau_0$ and
partons just stream freely between $\tau_0$ and $\tau_i$ (see details
for the $\tau$-evolution of the phase-space density in
Refs.~\cite{sin,Gyul}), then the system is suddenly thermalized at
$\tau_i$ with the pressure $P(\epsilon;\tau_i,x)$ corresponding to
the lattice QCD results. Then for $\tau_0=0.3$ fm/c, that
corresponds to $\Lambda_s=2~GeV$ \cite{KNV}, we find for the energy
density profile: $\epsilon(\tau_0,b=0;{\bf r}=0)\approx
0.09\Lambda_s^4/g^2$, Gaussian width $R(\tau_0, b=0)\approx 5.33$ fm
$(R_x=R_y=R)$. To describe pion spectra and interferometry radii one
should use \cite{QM}  $\Lambda_s^4/g^2\approx$ 6 GeV$^4$ and so, $
\epsilon(\tau_0=0.3$ fm/c, ${\bf r}=0)\equiv \epsilon_0= 67$
GeV/fm$^3$. For non-central collisions with $b=6.3$ fm: $R_x = 3.7$
fm, $R_y= 4.6$ fm.

The results for the collective flows are presented in  Fig. 1 (b=0
and b = 6.3 fm). We see qualitatively the same effect of the
transverse flow and their anisotropy in non-cental collisions at
pre-thermal stage as in the analytic non-relativistic examples.

\begin{figure}\label{fig1}
 \vspace{-0.3cm}
 \hspace{-3cm}
 \includegraphics[scale=0.85]{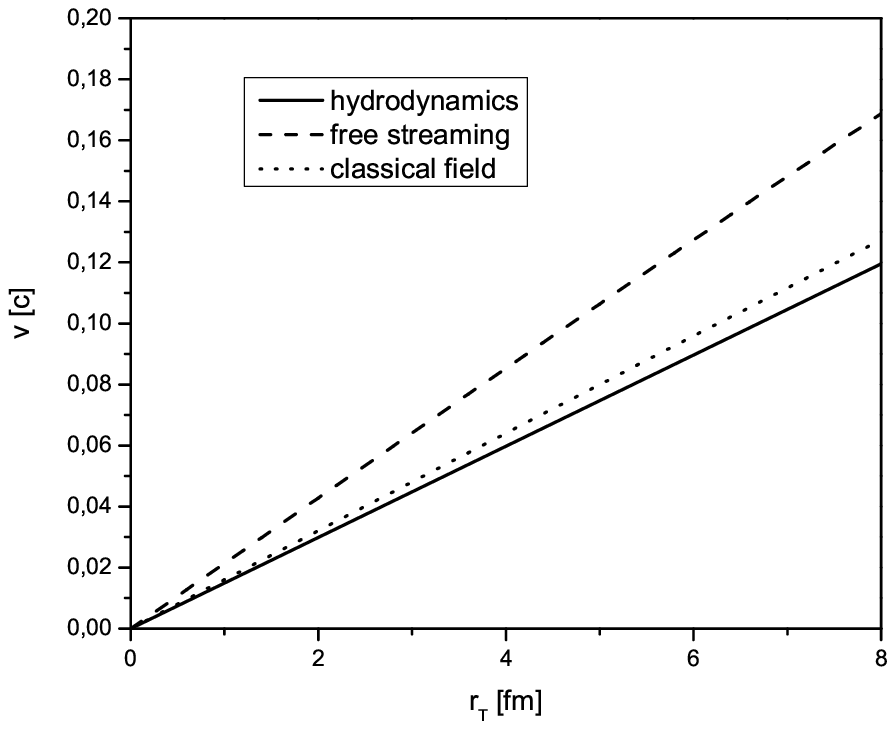}
 \includegraphics[scale=0.85]{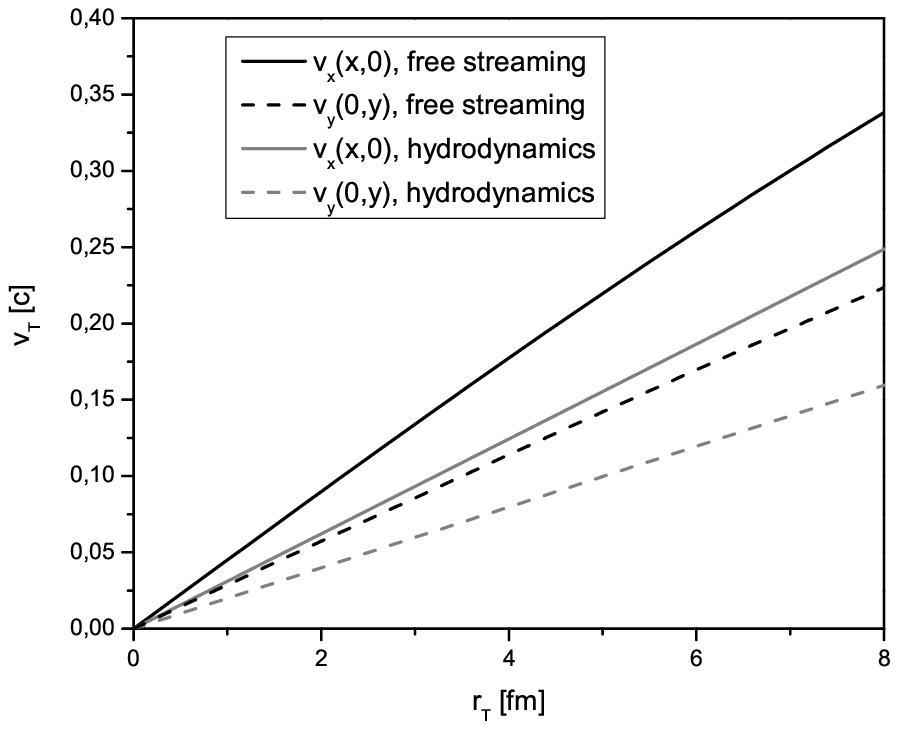}\hspace{-2cm}
 \vspace{-0.9cm}
 \caption{Collective velocity developed at
 pre-thermal stage from proper time $\tau_0=0.3$ fm/c by supposed thermalization time $\tau_i=1$ fm/c
 for scenarios of partonic free streaming and free expansion of classical  field. The results are compared with the hydrodynamic evolution of perfect fluid
 with hard equation of state $p=\frac{1}{3}\epsilon$ started at $\tau_0$. Left : central (b=0), right : non-central (b=6.3 fm) collisions.}
\end{figure}

\subsection{Field expansion}
Let us describe now the flows  at the early pre-thermal stage in
Glasma approach \cite{GV} which deals with evolution/rearrangement of
the classical gluonic field.  Let us again simplify the situation
and consider extreme case of the boost-invariant evolution of the
free field after moment $\tau_0\approx3/\Lambda_s$~\cite{KNV}. Then
we actually come to the Maxwell theory with 4-potential $A_\mu$ in
space-time with the pseudo-cylindrical metric
$ds^2=d\tau^2-\tau^2d\eta^2-dx^2-dy^2$.

Imposing the gauge $A_\tau=0$ used within the CGC concept, the
dynamical variables of the model are $\Phi\equiv A_\eta$, $A_i$
($i=x,y$). We will assume that these potentials and all the
observables do not depend on space-time rapidity $\eta$. It turns
out that $A_x$ and $A_y$ are related by condition $\partial_i A_i=0$
following from the gauge-fixing.

The space-time evolution of the potentials is determined by the
Maxwell equations which read
\begin{eqnarray}
&&\partial^2_\tau A_i+\frac{1}{\tau}\partial_\tau A_i-\Delta_T A_i=0,\\
&&\partial^2_\tau\Phi-\frac{1}{\tau}\partial_\tau\Phi-\Delta_T\Phi=0,
\end{eqnarray}
where $\Delta_T=\partial_i\partial_i$.

The components of the energy-momentum tensor, which are needed for
finding collective velocity, are defined by the well-known formula:
\begin{equation}
T^{\mu\nu}=-F^{\mu\lambda}F^\nu{}_\lambda+\frac{1}{4}g^{\mu\nu}F^{\lambda\sigma}F_{\lambda\sigma},\label{tensor}
\end{equation}
where $F_{\lambda\sigma}=\partial_\lambda A_\sigma-\partial_\sigma
A_\lambda$.

If $T^{\mu\nu}$ are known, the collective velocity is determined in
accordance with the Landau-Lifshitz definition from equations:
\begin{equation}
u^\mu=\frac{T^{\mu\nu}u_\nu}{u^\lambda
T_{\lambda\sigma}u^\sigma},\quad u^2=1.\label{flow}
\end{equation}

Now let us focus on the initial conditions for field
equations. We apply the Glasma-like ones~\cite{GV}:
\begin{eqnarray}
&&\partial_\tau A_i|_{\tau_0}=0,\quad \Phi|_{\tau_0}=0,\\
&&E_z\sim
\left.\frac{\partial_\tau\Phi}{\tau}\right|_{\tau_0}\not=0,\quad
H_z\sim F_{xy}|_{\tau_0}\not=0.
\end{eqnarray}
These equations
mean that, at the initial moment $\tau_0$, the
electric and magnetic fields are longitudinal, while the transverse
components vanish. So there is no field flow at $\tau_0$.

With those initial conditions we get (at $\eta=0$)
\begin{eqnarray}
T_{tt}|_{\tau_0}&\equiv&\varepsilon(r_T)=
\frac{1}{2}\left(\left.\frac{\partial_{r_T}\Psi}{r_T}\right|_{\tau_0}\right)^2
+\frac{1}{2}\left(\left.\frac{\partial_\tau\Phi}{\tau}\right|_{\tau_0}\right)^2,\label{e}\\
T_{tx}|_{\tau_0}&=&0.
\end{eqnarray}

Let us choose the initial energy density profile
$\varepsilon(r_T)=\epsilon_0\exp{\left(-\frac{r^2_T}{2R^2}\right)}$
to be the same as in the previous case of partonic system for b=0.

To provide this we use, accounting for (\ref{e}), the following
transverse profile for potentials at the initial proper time
$\tau_0$:
\begin{equation}
\partial_{r_T}\Psi|_{\tau_0}=\sqrt{\alpha}r_Tf(r_T),\quad
\partial_\tau\Phi|_{\tau_0}=\sqrt{1-\alpha}\tau_0f(r_T),
\end{equation}
where $f(r_T)\equiv\sqrt{2\varepsilon(r_T)}$ and $\alpha$ is a
constant. Since the potentials $\Psi$, $\Phi$ are real, $\alpha$ bounds are
$0\leq\alpha\leq1$. The observable quantities
$T_{tt}(\tau,r_T,\varphi=0,\eta=0)$,
$T_{tx}(\tau,r_T,\varphi=0,\eta=0)$, and $v(\tau,r_T)$ are
independent on parameter $\alpha$.

Further, we will treat these relations (together with
$\partial_\tau\Psi|_{\tau_0}=0$ and $\Phi|_{\tau_0}=0$) as the
initial conditions for our problem.

The solution for the field is expressed through the field amplitude
at the initial moment $\tau_0$:
\begin{equation}
\Psi_0(k_T)=\sqrt{\alpha}\tilde f(k_T),\quad
\Phi_0(k_T)=\sqrt{1-\alpha}\tilde f(k_T),
\end{equation}
here
\begin{equation}
\tilde f(k_T)=\int\limits_0^\infty f(r_T)J_0(k_Tr_T)r_Tdr_T.
\end{equation}

Then the time-dependence of amplitudes takes the form
\begin{eqnarray}
&&\tilde\Psi(\tau,k_T)=\Psi_0(k_T)\frac{\pi k_T\tau_0}{2}
[J_1(k_T\tau_0)Y_0(k_T\tau)-J_0(k_T\tau)Y_1(k_T\tau_0)],\\
&&\tilde\Phi(\tau,k_T)=\Phi_0(k_T)\frac{\pi k_T\tau}{2}
[J_1(k_T\tau_0)Y_1(k_T\tau)-J_1(k_T\tau)Y_1(k_T\tau_0)].
\end{eqnarray}

To restore the spatial dependence, the Bessel-Fourier transform
should be applied:
\begin{eqnarray}
&&\Psi(\tau,r_T)=r_T\int\limits_0^\infty\tilde\Psi(\tau,k_T)J_1(k_Tr_T)dk_T,\\
&&\Phi(\tau,r_T)=\int\limits_0^\infty\tilde\Phi(\tau,k_T)J_0(k_Tr_T)dk_T,
\end{eqnarray}
where an integration over the angle has been already carried out. It
allows one to calculate the evolution of  the momentum-energy tensor
(\ref{tensor}) and then find the transverse flows according to Eq.
(\ref{flow}). The results are presented at the Fig. 1. One  can see
that they are similar to ones obtained for partonic free streaming.

\subsection{Hydrodynamic evolution}
 We compare the developing of the transverse collective velocities at pre-thermal stage
 by the time $\tau_i$= 1 fm/c with the hydrodynamic evolution of perfect fluid
 having the hard equation of state (EoS) $p=\frac{1}{3}\epsilon$.
 The initial conditions for hydro expansion: the time $\tau_0$ and the energy density profile are supposed to be the same as
 in previous cases (subsections 2 and 3).
 One can see from Fig. 1 that the
hydrodynamic expansion is relatively less effective for producing
transverse flows and their anisotropy than the 
non-interacting evolution.

\section{Conclusion}
We considered a possible developing of the transverse flows in
different extreme cases: the evolution of non-relativistic free
gas, partonic free streaming, classical field expansion and
hydrodynamic expansion of the perfect fluid with the hard equation
of state. All the results are similar, though in relativistic
situation the transverse flows, as well as their anisotropy, are
developed at pre-thermal, either partonic or classical field,
stage with even more efficiency than in the case of very early
perfect hydrodynamics with hard EoS. So, one  can conclude that
the transverse flows and their anisotropy in non-central A+A
collisions, in fact, do not depend on the nature of the systems
formed and complicated kinetics of non-equilibrium processes. They
are defined mostly by the homogeneity lengths in the initial
system. However, that is crucially important, the transformation
of these radial and elliptic flows into observed high effective
temperatures of proton and kaon spectra and anisotropy of pion
spectra becomes possible {\it only if (local) thermalization
happens}. Otherwise, e.g., at a free streaming, the initial
partonic momentum distribution and its symmetry properties are
preserved despite the flow and its asymmetry in non-central
collisions are developed. So, the thermalization is certainly
required for description of the hadronic spectra and $v_2$
coefficients  but it does not matter whether thermalization is
early or not.

\section*{Acknowledgments}
This work was supported in part by the Fundamental Research State
Fund of Ukraine, Agreement No. F25/239-2008; the Bilateral award DLR
(Germany) - MESU (Ukraine) for the UKR 06/008 Project, Agreement No.
M/26-2008; and the Program ``Fundamental Properties of Physical
Systems under Extreme Conditions"  of the Bureau of the Section of
Physics and Astronomy of NAS of Ukraine. The research was carried
out within the scope of the ERG (GDRE): Heavy ions at
ultrarelativistic energies -— a European Research Group comprising
IN2P3/CNRS, Ecole des Mines de Nantes, Universit\'e
de Nantes, Warsaw University of Technology, JINR Dubna, ITEP Moscow,
and Bogolyubov Institute for Theoretical Physics, NAS of Ukraine.


\begin{thebibliography}{99}

\bibitem{Landau} L.D. Landau, Izv. Akad. Nauk SSSR, Ser. Fiz.
{\bf 17}, 51 (1953);U. Heinz, J. Phys. G: Nucl. Part. Phys. {\bf
31}, S717 (2005); T. Hirano, Nucl. Phys. A {\bf 774}, 531 (2006).

\bibitem{hybrid} S.A. Bass, A. Dumitru, Phys. Rev C {\bf 61}, 064909
(2000).

 \bibitem{PRC}S.V. Akkelin, Y. Hama, Iu.A. Karpenko, and Yu.M.
Sinyukov, Phys. Rev. C {\bf 78}, 034906 (2008).

 \bibitem{STAR} K.H. Ackermann et al. (STAR Collaboration), Phys. Rev. Lett.
 {\bf 86},402 (2001).

\bibitem{elliptic} P.F. Kolb, P. Huovinen, U. Heinz, H. Heiselberg, Phys.Lett. B {\bf 500}, 232
(2001); U. Heinz, P. Kolb Nucl. Phys. A {\bf 702}, 269 (2002).

\bibitem{Borysova} M.S. Borysova, Yu.M. Sinyukov, S.V. Akkelin, B. Erazmus,
and I.A. Karpenko, Phys. Rev. C {\bf 73}, 024903 (2006).

\bibitem{sin}~Yu.M. Sinyukov,  \textit{Act.Phys.Polon.\/} B {\bf 37}, 3343 (2006).

\bibitem{early} Z. Xu, C. Greiner and H. Stocker J. Phys. G: Nucl. Part. Phys. {\bf 35} 104016
(2008).

\bibitem{Gyul}M. Gyulassy,~Iu.A. Karpenko,~A.V. Nazarenko,~Yu.M. Sinyukov
\textit{Braz. Journ. of Phys.\/}, {\bf 37}, 1031 (2007).

\bibitem{LHC} Yu.M. Sinyukov,~S.V. Akkelin,~Iu.A. Karpenko, \textit {Phys. Atomic Nuclei}, {\bf 71} (2008)
1619; \textit{arXiv: 0706.4066\/}; N. Armesto et al,  J. Phys. G:
Nucl. Part. Phys. {\bf 35},  054001 (2008).

\bibitem{QM} Yu.M. Sinyukov, Iu.A. Karpenko, A.V. Nazarenko, J. Phys. G: Nucl. Part. Phys. {\bf 35}, 104071 (2008).

\bibitem{Florkowsky}W. Broniowski,M. Chojnacki, W. Florkowski, A.
Kisiel, arXiv:0801.436 (2008).

\bibitem{Pratt}J. Vredevoogd, S. Pratt, arXiv:0810.4325 (2008).

\bibitem{AkkSin} S.V. Akkelin, Yu.M. Sinyukov, Phys. Rev. C {\bf 70} , 064901
(2004).

\bibitem{KNV}A. Krasnitz, R. Venugopalan \textit{Phys. Rev. lett.\/}
{\bf 84} (2000) 4309; A. Krasnitz, Y. Nara, R. Venugopalan:
\textit{Nucl. Phys. A {\bf 717}\/}, 268  (2003); A {\bf 727}, 427
(2003).

\bibitem{GV}
T. Lappi, L. McLerran, Nucl. Phys. A {\bf 772}, 200 (2006);\\
 F. Gelis
and R. Venugopalan, Acta Phys. Polon. B {\bf 37}, 3253 (2006).
\end{thebibliography}
\end{document}